**Star Watch Astrometry Probe**
**July 10, 2019**
**Philip Horzempa**

The technology is now in hand to detect Earth Analogs in nearby star systems. This paper describes the Star Watch astrometry spacecraft. It utilizes a Michelson interferometer to achieve uas-level stellar measurement precision. The heart of this detector is the Astrometric Beam Combiner. Remarkably, this hardware is in storage at JPL, ready to complete testing, followed by an upgrade to flight status.

Star Watch is a space borne instrument, capable of astrometric measurements to micro-arc second precision for a large sample of stars in our galaxy. It builds on the technology developed during the SIM (Space Interferometery Mission) and will carry out a 5 year mission from an Earth-Trailing Solar Orbit. The single optical instrument is a stellar optical interferometer system with 50cm collecting apertures separated by a 6 meter baseline. It includes one "guide" interferometer and one "guide" telescope for spacecraft pointing reference and one "science" interferometer to perform high accuracy astrometric measurements on target stars.

There is essentially no technology development required for the Star Watch Interferometer instrument. A multi-year program of design, test, and risk retirement under the auspices of the SIM (Space Interferometry Mission) program produced hardware that was almost flight-ready.

During its lifetime, the SIM project successfully addressed many technological challenges to show that the mission was technically feasible. These challenges range from nanometer-level control to picometer-level sensing. Key testbeds and brass-board components were designed, built, and tested during the development phase of SIM, resolving all the major technology challenges. They confirmed that it would be able to make astrometric observations at the level of 0.1-0.5 uas.
Star Watch will build on that invaluable engineering heritage, incorporating advances in the state-of-the-art since 2010. These include the development of smaller, lighter beam launchers and corner cubes for laser metrology, attitude-control micro-thrusters that eliminate disturbances from reaction wheels, and advanced fringe detectors.

**Science Motivations**

There is a now an impasse in exoplanet research. Transit and Radial Velocity searches have performed remarkable work in discovering 2,000 exoplanets in the past two decades. These objects come in a variety of sizes and orbits, but physical limitations have prevented these search efforts from detecting Earth Analogs. These are defined as planets of 1 Me, orbiting a G star, in an orbit with a semi-major axis of 1 AU. Star Watch will detect those worlds around stars that lie within 20 parsecs of our solar system.

Star Watch addresses one of the key recommendations of a NASA's Exoplanet Science Strategy. It will perform a "survey for planets where the census is most incomplete, which includes the

parameter space occupied by most planets in the Solar System."

   Besides the goal of Earth Analog detection, Star Watch will allow, for the first time, detection of sub-Jovian planets in orbits with semimajor axes of 2 AU, or larger.  For example, a population of Super Earths has been proposed to constitute an "Invisible Majority" of worlds located in the outer reaches of star systems.  With masses of 5 Me and semimajor axes of 5 AU, only uas-level astrometry will be able to verify, or refute, this hypothesis.  Either judgment will allow theoretical work to proceed on a solid footing of data.
   As a bonus, Star Watch will characterize the planetary architecture of some 1,000 stars by looking for planets in the range from Super Earths to Jupiter.

   While Star Watch can only deliver accurate at-epoch positions on the sky in the most favorable cases, in all cases it provides mass, orbit semi-major axis and orbital inclination, all required to confirm the discovery of a terrestrial, habitable zone planet.   This is the existence proof – that a star has an Earthlike (terrestrial, habitable-zone) planet (or that it does not), critical information for a follow-on mission whose main goal is spectroscopy, not discovery.

**"Angle of the Light"**

   The Star Watch Astrometric Observatory is designed to perform just one kind of measurement – precision astrometry – that enables its entire science program.    Each science objective is achieved by measuring the astrometric signature of the object under study.

   It utilizes a Michelson stellar interferometer operating in the visible spectrum. It consists of two 50 cm siderostats separated by the 6 meter baseline. Light reflecting from the two siderostats is collected by telescopes and propagated by a set of optics to the astrometric beam combiner (ABC) where the two optical wavefronts are recombined, forming interference fringes. The peak interference fringe is obtained when the propagation path through the two arms of the instrument is identical.

   The basic elements of a stellar interferometer are shown in Figure 1. Light from a distant source is collected at two points and combined using a beam splitter, where interference of the combined optical wavefronts produces fringes when the internal optical path-length difference (OPD, or delay,) compensates exactly for the external delay. The astrometric angle α between the interferometer baseline and the ray from the star can be measured if the length of the baseline $B$ and the internal delay $x$ are measured.

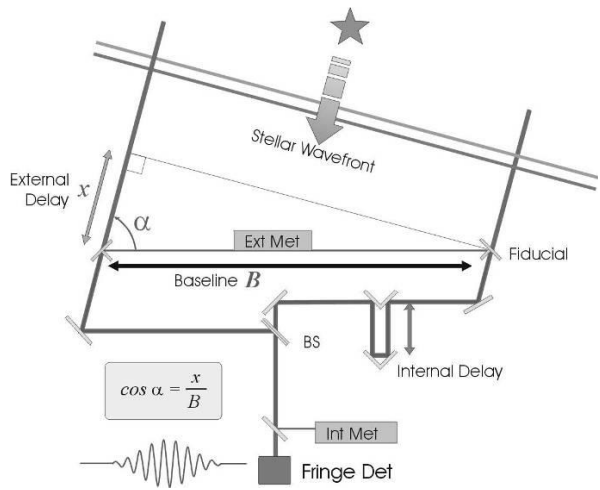

**Figure 1. Essentials of Star Watch interferometry.**

For Star Watch, the baseline is defined by two fiducials, each made of common-vertex corner cubes and located at the center of the starlight collecting apertures. The external metrology system measures $B$, the distance between the two fiducials and the internal metrology system measures $x$, the optical path difference to the beam combiner from the same two fiducials. The starlight fringe detector measures any residual optical path difference through the two arms of the interferometers.

The science interferometer makes measurements of the path delay (through the instrument) of stars observed sequentially. These measurements can be processed to represent angles on the sky projected along the interferometer baseline.

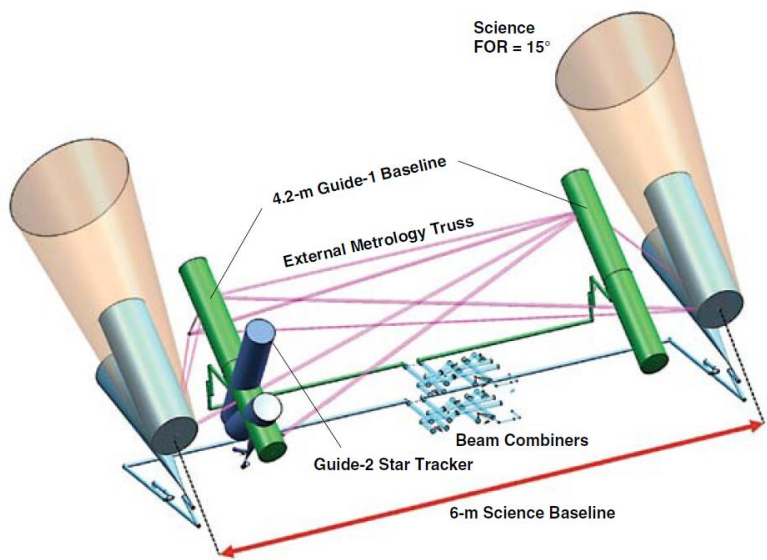

**Figure 2: Essentials of Star Watch Optical Bench**

The precision metrology systems only measure pathlength changes; therefore Star Watch makes differential angle measurements between stars. The overall delay and hence the overall angle are not measured. Similarly, it is not the absolute baseline vector that is measured by the external metrology system, but the changes in the baseline vector length and orientation during the observations.

In between each observation, the two siderostats and the optical delay line have to be repositioned to acquire fringes on the next star to observe. Once all the reference stars have been observed, the observation sequence is repeated from the beginning. After repeated visits to the target star, the astrometric noise will be below 0.035 µas, enabling detection of astrometric signatures of 0.2 µas with a signal to noise ratio of 6. As a reference, the signature of the Earth is 0.3 µas for an observer located 10 parsecs away.

The required performance of 24 and 15 picometers for the interferometer, 50 micro-arc-seconds for the guide telescope and 6 picometers for the external metrology has been experimentally demonstrated on testbeds built during the SIM project.

**Engineering Risk Reduction**

After the completion of SIM's technology program in 2005, the next step was construction of flight-grade hardware. Between 2005 and 2006, brass-board models of the internal metrology launcher, external metrology launcher, metrology fiducials and laser metrology source bench were tested and qualified
These brassboards (form, fit and function to flight) of key instrument assemblies were subjected to full flight qualification-level environments and rigorous pre/post environmental performance testing. This allows proceeding directly to protoflight units.

A brass-board of the 30 cm optical beam compressor and Siderostat mirror were also built for thermal stability testing, which was successfully completed. The next engineering step was the demonstration of the ability to mount the large precision optics without distorting the reflected wavefront, while being able to survive launch. This test was successfully completed with the optical compressor 34 cm primary mirror.

In 2007, the Guide 2 Telescope (G2T) testbed, a demonstration of the Guide 2 Telescope concept, was initiated to successfully validate the Guide 2 error budget. At that time, the Sim project built brass-board models of all the key components of the interferometer: Astrometric Beam Combiner, focal-plane detectors with electronics and cryo-heat-pipe cooling, precision pointing and phasing mechanisms.

The Instrument System and Sub-systems are rated at TRL6. The technology required to perform the science has been demonstrated in the relevant vacuum environment, with thermal and mechanical stability conditions worse than those expected on orbit, using a suite of hardware and software testbed demonstrations.

The SIM project demonstrated all of the technology and engineering needed for the flight instrument. Every component was vetted for manufacturing, technology and performance risks prior to the end of FY2010. Star Watch is technically ready for full-scale development.

 **List of technology developments**

 During the SIM project, all subsystem components were developed to TRL 6 or higher. In some subsystems, further performance improvements are expected prior to PDR. For example, the dual stage reaction wheel isolator will be replaced by micro-thrusters. This removes a major source of instrument noise. In addition, light-weight laser beam launchers have been developed by the communications industry since 2010, allowing the replacement of the bulkier units that were part of the SIM-Lite design.

 Since 2010, the GRACE-Follow On and LISA Pathfinder missions have successfully flown laser metrology systems.

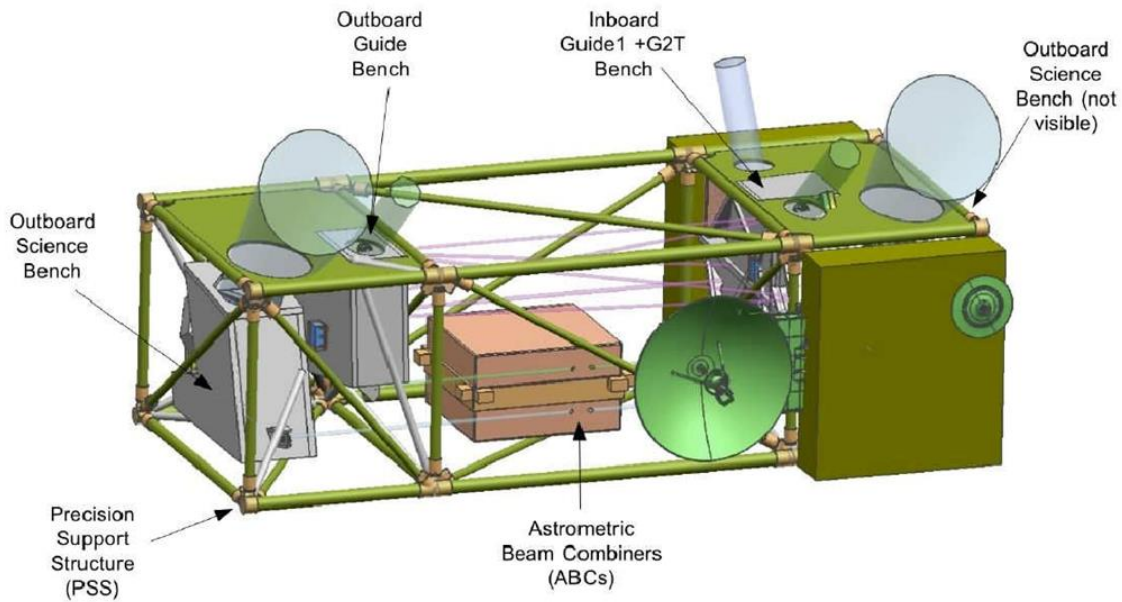

**Figure 3: Star Watch spacecraft structure**

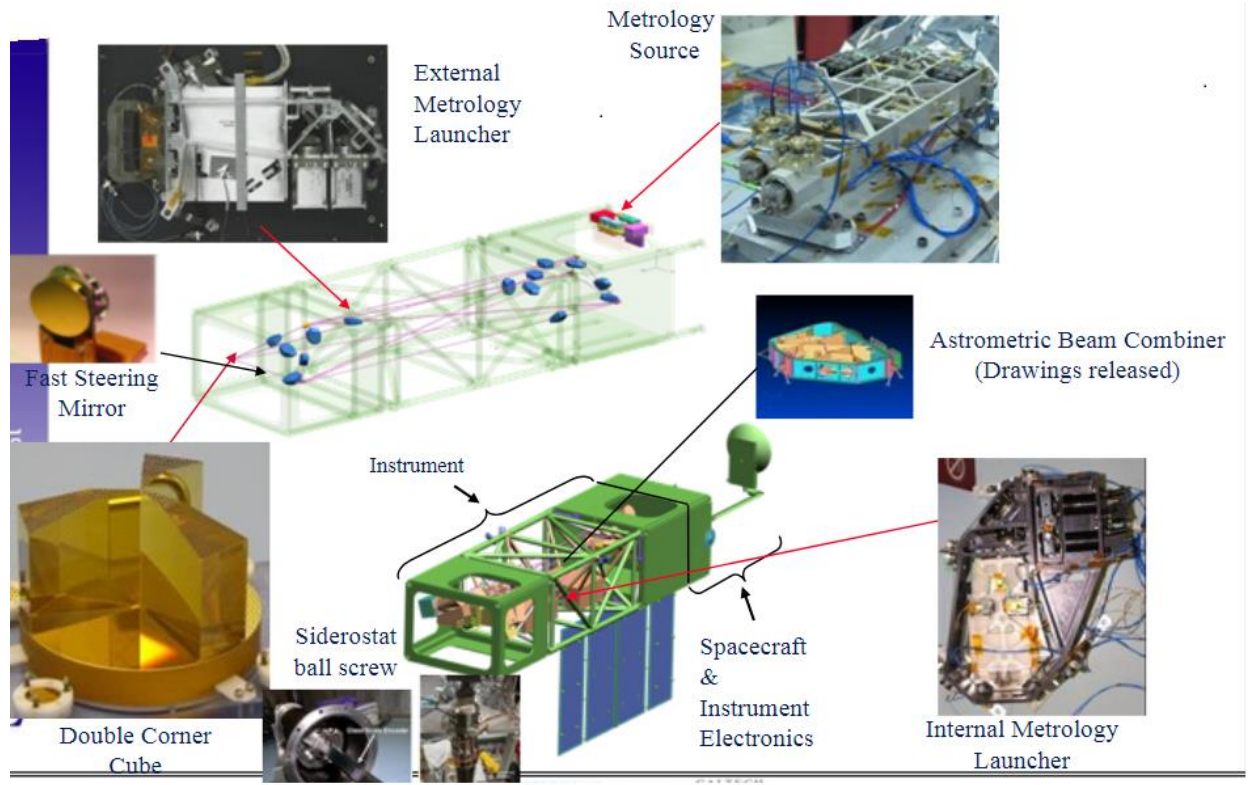

**Figure 4. Examples of key hardware.**

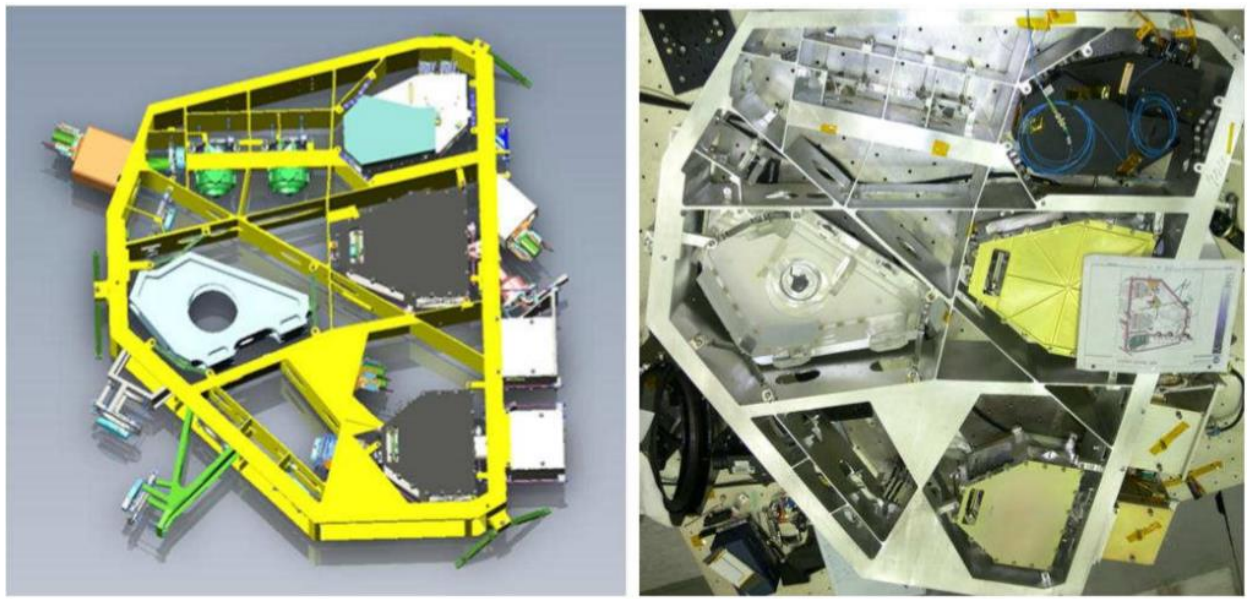

**Figure 5. (left) Illustration of physical arrangement of the ABC.
 (right) Photo of ABC hardware.**

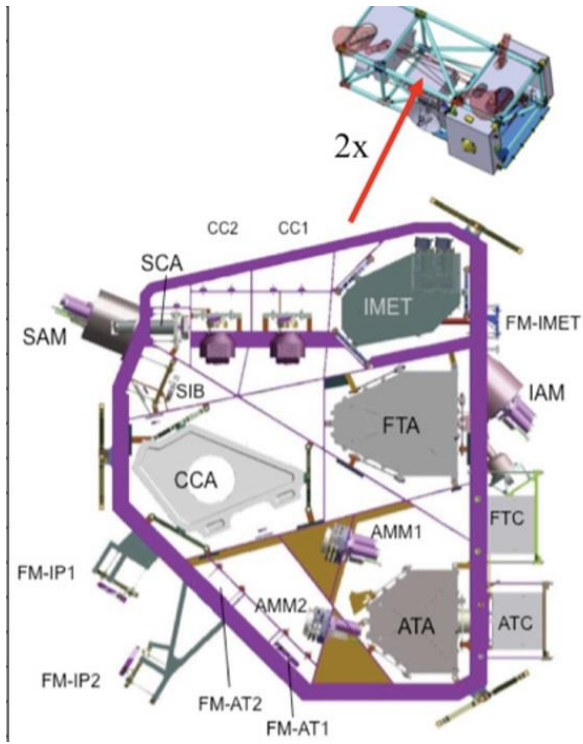

**Figure 6:** Some of the components of the ABC are shown. These include the FTC and FTA, which are the Fringe Tracking Camera and Assembly. The ATA and ATC are the Angle Tracking Camera and Assembly.

**Path Forward**

With the key optical metrology hardware built and tested, Star Watch can begin an expedited transition into Phase C. The first step will be to remove the ABC from bonded storage. It can then be integrated into a new BII&T testbed that incorporates modern optical and electronic technology.

An RFP will be issued as the ABC is being prepared. This request will ask for bids for payload integration and development of a spacecraft bus.

**Technology Driver for the Future**

The technology pioneered by Star Watch, the first long-baseline Michelson Interferometer in space, represents an important investment for the future of space astronomy. Picometer-level laser metrology developed for Star Watch is required for future segmented-mirror large space telescopes.

In addition, all of NASA's proposed Vision Missions, such as the Exo-Earth Mapper, Black Hole Mapper and Cosmic Dawn Mapper, require the use of precision interferometry. Star Watch will provide a path to those missions.

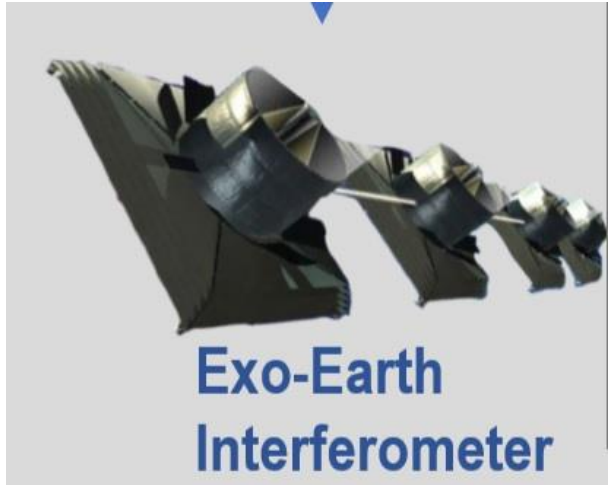

**Figure 7: Exo Earth Interferometer**

One could envision a nano-arcsecond astrometry mission as a follow-on to Star Watch. This would allow the reach of this method to extend to 50-100 parsecs. In addition, the proposed mission to the Sun's Gravity Lens Focus requires determination of an exoplanet astrometric orbit at ~10 nas.

**Conclusions**

No other Astrophysics Probe mission concept comes close to this level of technical readiness to proceed to ATLO. There are no analogs to Star Watch. It is unique and is a natural next step in the effort to detect, and characterize, exoplanets. Star Watch will enable access to the realm of temperate Terrestrial worlds for the first time, measuring their masses and orbits. This can be achieved by the middle of the 2020s, i.e., 20-25 years before Direct Imaging by space telescopes.
  In the search for, and survey of, other worlds, it is incumbent on the exoplanet community to bring all tools to bear. The science measurements that Star Watch will provide cannot be duplicated by other means.

Star Watch is a mission whose time has come. It will provide data to confirm RV planets and will detect the "golden targets" for future Direct Imaging missions. It will provide crucial velocity and mass data for the fields of Dark Matter research and stellar evolution.
  The march of progress makes this space observatory lighter and more capable than SIM.

Geometry and Physics have conspired to keep nearby Earth Analogs hidden. We expect that they exist. Transit detection has only a vanishingly small of detection because of orbital alignment requirements. Radial Velocity efforts are hindered by a 1 m/sec "Doppler Wall" of spectral noise caused by stellar jitter. The means exist to overcome these barriers through the use of uas-level astrometry. Star Watch is the vehicle by which that capability can be exercised. The Decadal Survey is asked to endorse this concept.

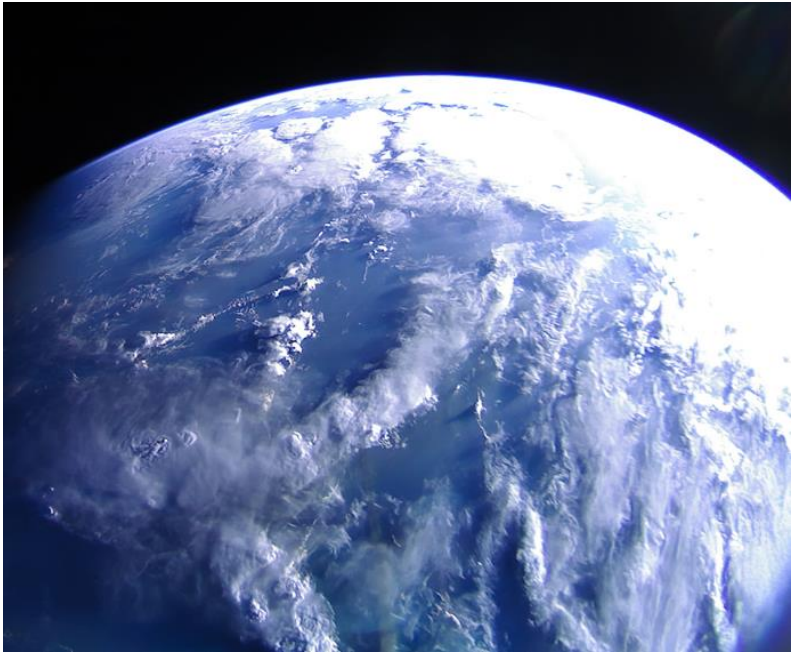

**Figure 8: Objective**

**References**

**Recent Status of SIM Lite Astrometric Observatory Mission: Flight Engineering Risk Reduction Activities**
 **Goullioud et al; 2010**

**The Invisible Majority? Evolution and Detection of Outer Planetary Systems without Gas Giants**
**Mann et al; 2010**

**Synthesis Imaging at Optical Wavelengths with SIM**
**Allen, R.; 2005**